\documentclass[reprint,amsmath,amssymb,aps,prl]{revtex4-1}

\def\sr/{$^{88}$Sr$^{+}$}
\def\pone/{$5P_{1/2}$}
\def\pthree/{$5P_{3/2}$}
\def\sone/{$5S_{1/2}$}
\def\dthree/{$4D_{3/2}$}
\def\dfive/{$4D_{5/2}$}

\def\p/{17.175(27)}
\def\q/{15.845(71)}
\def\s/{0.05609(21)}
\def\rrr/{0.0063(4)}

\usepackage{amssymb}
\usepackage{multirow}
\usepackage{tabularx}
\usepackage{graphicx}

\usepackage{hyperref}
\hypersetup{
   bookmarksnumbered=false,
   unicode=true,
   pdfstartview={FitH},
   pdfnewwindow=true,
   colorlinks=true, 
   linkcolor=blue,
   citecolor=blue,
   filecolor=blue, 
   urlcolor=blue 
}

\begin{document}

\title{Iterative Precision Measurement of Branching Ratios Applied to 5$P$ states in $^{88}$Sr$^+$}

\author{Helena Zhang}
\thanks{These two authors contributed equally to this work.}
\author{Michael Gutierrez}
\thanks{These two authors contributed equally to this work.}
\author{Guang Hao Low}
\author{Richard Rines}
\author{Jules Stuart}
\author{Tailin Wu}
\author{Isaac Chuang}
\email{ichuang@mit.edu}
\affiliation{Massachusetts Institute of Technology, Cambridge, Massachusetts 02139, USA}

\begin{abstract}
We report on a method for measuring the branching ratios of dipole transitions of trapped atomic ions by performing nested sequences of population inversions. This scheme is broadly applicable and does not use ultrafast pulsed or narrow linewidth lasers. It is simple to perform and insensitive to experimental variables such as laser and magnetic field noise as well as ion heating. To demonstrate its effectiveness, we make the most accurate measurements thus far of the branching ratios of both \pone/ and \pthree/ states in \sr/ with sub-1\% uncertainties. We measure \p/ for the branching ratio of \pone/--\sone/, \q/ for \pthree/--\sone/, and \s/ for \pthree/--\dfive/, ten-fold and thirty-fold improvements in precision for \pone/ and \pthree/ branching ratios respectively over the best previous experimental values.
\end{abstract}

\maketitle

Empirical measurements of elemental constants are fundamental to the verification and advancement of our knowledge of atoms. One important atomic property is the branching ratio of an electron transition: the ratio of its transition rate to the sum of rates of other decay channels with the same excited state. Measuring these constants accurately is vital for the refinement of relativistic many-body theories and provides a crucial probe in the study of fundamental physics such as parity non-conservation \cite{safronova2010all,jiang2009blackbody,guet1991relativistic,fortson1993possibility,dzuba2011calculation,dutta2014}.

Branching ratios for different atomic species are also of great use in a wide range of fields including astrophysics, where analyzing the composition of stars contributes greatly to understanding stellar formation and evolution. Abundances of heavy elements such as strontium are essential for determining the efficiency of neutron capture processes in metal-poor stars, yet can be difficult to determine from emission spectra due to nearby transitions of other elements  \cite{caffau2016giano,hansen2014exploring,brage1998theoretical,bautista2002excitation,gratton1994abundances,siqueira2015high,bergemann2012nlte}. Branching ratios of these transitions are therefore vital for quantitative modeling of nucleosynthesis processes \cite{bergemann2012nlte,siqueira2015high,gratton1994abundances}.

In addition, precise branching ratios enable the improvement of clock standards, paving the way for better global positioning systems and tests of the time-invariance of fundamental constants \cite{ludlow2015optical}. Atomic clocks using the optical quadrupole transition \sone/--\dfive/ in \sr/, one of the secondary clock standards recommended by the International Committee for Weights and Measures \cite{ludlow2015optical}, have achieved uncertainties at the $10^{-17}$ level \cite{madej2012sr+}, more accurate than the current $^{133}$Cs clock standard \cite{weyers2011distributed}. To further improve the precision of these systems, it is necessary to reduce uncertainty from the blackbody radiation Stark shift, the dominant source of error in room temperature clocks \cite{jiang2009blackbody}. Branching ratios measured at the 1\% level, combined with high-precision lifetime measurements, would allow for a significant reduction in blackbody radiation shift error by improving the accuracy of static polarizabilities of clock states \cite{jiang2009blackbody,madej2012sr+}.

Despite their relevance, branching ratios of heavy atoms have not been precisely measured for many decades due to the large uncertainties inherent in traditional discharge chamber methods using the Hanle effect \cite{gallagher1967oscillator}. Recent astrophysical studies still use these older experimental results for fitting emission spectra \cite{siqueira2015high,bergemann2012nlte}. Only in the last decade have there been precision measurements of branching ratios at the 1\% level \cite{ramm2013precision,likforman2015precision,gerritsma2008precision,de2015precision} using trapped ions, versatile toolkits for spectroscopy \cite{hettrich2015measurement,madej2012sr+,fortson1993possibility} as well as quantum computation \cite{monz2015realization}. In particular, pioneering work has been done by Ramm et. al. \cite{ramm2013precision}, establishing benchmark results for $^{40}$Ca$^{+}$ and methods for three level lambda transition systems.

Here, we present a novel scheme for measuring the branching ratio of the $P_{3/2}$ state of a trapped ion with an iterative population transfer sequence, building upon this prior art. As with \cite{ramm2013precision}, we do not require ultrafast pulsed lasers or narrow linewidth lasers for addressing quadrupole transitions, which were used by previous precision measurements of $P_{3/2}$ branching ratios \cite{kurz2008measurement,gerritsma2008precision}. Our method uses only two lasers that pump the ion from the ground state to the $P_{1/2}$ and $P_{3/2}$ excited states and two lasers to unshelve the ion from the metastable states below $P_{3/2}$. For \sr/ and analogous species, these lasers are already used for Doppler cooling, making this scheme broadly applicable for many trapped ion systems without the need for additional equipment. Like \cite{ramm2013precision}, our method is insensitive to experimental variables such as magnetic field and laser fluctuations, but what we present extends beyond lambda systems to allow branching ratios of more complex systems to be obtained. We demonstrate the effectiveness of our method by making the first precision measurement of the \pthree/ branching ratios in \sr/ in addition to the most accurate measurement of the \pone/ branching ratios to date.

\begin{figure}
\includegraphics[width=8cm]{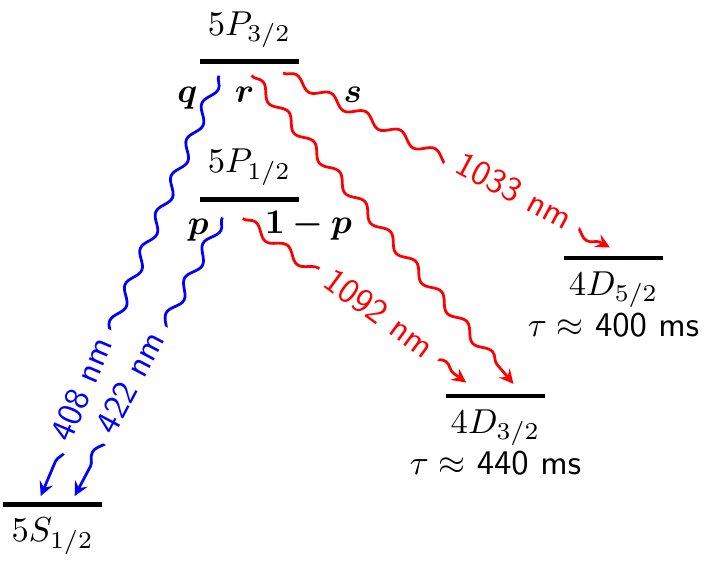}
\caption{Energy levels of \sr/, showing the \pone/ and \pthree/ excited states and their decay channels with transition wavelengths. \dthree/ and \dfive/ are metastable states with lifetimes much greater than the timescale of the experiment \cite{sahoo2006lifetimes}. To find the \pthree/ branching ratio, we need only lasers addressing the 408, 422, 1092, and 1033 nm transitions.}
\label{fig:levels}
\end{figure}

We begin by briefly describing the procedure for measuring branching ratios of $J=1/2$ states using the method by Ramm et. al, which will be a building block for the $J=3/2$ system. We use the \pone/ excited state in $^{88}$Sr$^+$ as the model system (Fig. \ref{fig:levels}). We denote the probability of decaying to the ground \sone/ state as $p$ and the long-lived \dthree/ state as $1-p$.

At the start of the experiment, the ion is intialized to the ground \sone/ state. In the first step, the 422 nm laser is turned on to invert population to the excited \pone/ state while we record ion fluorescence at 422 nm. As the ion decays to the metastable \dthree/ state, we detect a mean number of photons $\langle n \rangle=\epsilon_{422}\cdot p/(1-p)$, where $\epsilon_{422}$ is the detection efficiency of our system at 422 nm, before the ion is fully shelved. In the second step, the 1092 nm laser is turned on to repump the ion to the excited state, during which we detect $\epsilon_{422}$ photons as it decays to the \sone/ state. The branching ratio $p/(1-p)$ is therefore equal to the ratio of the number of counts observed during the two time intervals, independent of the collection efficiency.

For the more complex \pthree/ state, which decays to three instead of two states, we denote the probability of decaying to the \sone/, \dthree/, and \dfive/ states as $q$, $r$, and $s=1-q-r$ respectively (Fig. \ref{fig:levels}). To measure the \pthree/ branching ratios $q/(1-q)$, $r/(1-r)$, and $s/(1-s)$, we begin with a sequence analogous to the \pone/ sequence, this time detecting photons at both 408 nm and 422 nm. Starting again with the ion in the \sone/ ground state, we first pump the ion into the excited \pthree/ state with the 408 nm laser (Step A). We detect a mean number of photons from the ion
\begin{align}
\langle N_A \rangle= \displaystyle \epsilon_{408}\frac{q}{1-q},
\end{align}
where $\epsilon_{408}$ is the detection efficiency at 408 nm. We now turn on the 1033 nm laser, which drives the ion to the \pthree/ state if it was in the \dfive/ state and does nothing otherwise (Step B). We detect a mean number of 408 nm photons
\begin{align}
\langle N_B \rangle=\displaystyle \epsilon_{408}\frac{qs}{(1-q)(1-s)}
\end{align}
in this step. We can obtain the \pthree/--\dfive/ branching ratio $s/(1-s)$ from the photon count ratio of the previous two steps.

To measure the other two branching ratios, we note that their values are contained in the state of the ion after Step B---the population split between the \sone/ and \dthree/ states. To obtain this information, we now turn on the 422 nm laser to pump all \sone/ population into the \dthree/ state (Step C). We detect
\begin{align}
\langle N_C \rangle=\displaystyle \epsilon_{422}\frac{qs}{(1-q)(1-s)}\frac{p}{1-p}
\end{align}
photons at 422 nm. Finally, turning on the 1092 nm laser repumps all of the population to the \sone/ state and we detect $\epsilon_{422}$ photons (Step D), which is necessary for canceling the detection efficiency $\epsilon_{422}$.

Since we can determine $p$ experimentally with the \pone/ branching ratio sequence, we can solve for the \pthree/ branching ratios without knowing $\epsilon_{422}$ or $\epsilon_{408}$:
\begin{align}
\frac{s}{1-s}&=\frac{\langle N_B \rangle}{\langle N_A \rangle}\\
\frac{q}{1-q}&=\frac{\langle N_A\rangle\langle N_C\rangle}{\langle N_B\rangle\langle N_D\rangle}\frac{1-p}{p}
\end{align}

\begin{figure*}
\includegraphics[width=\linewidth]{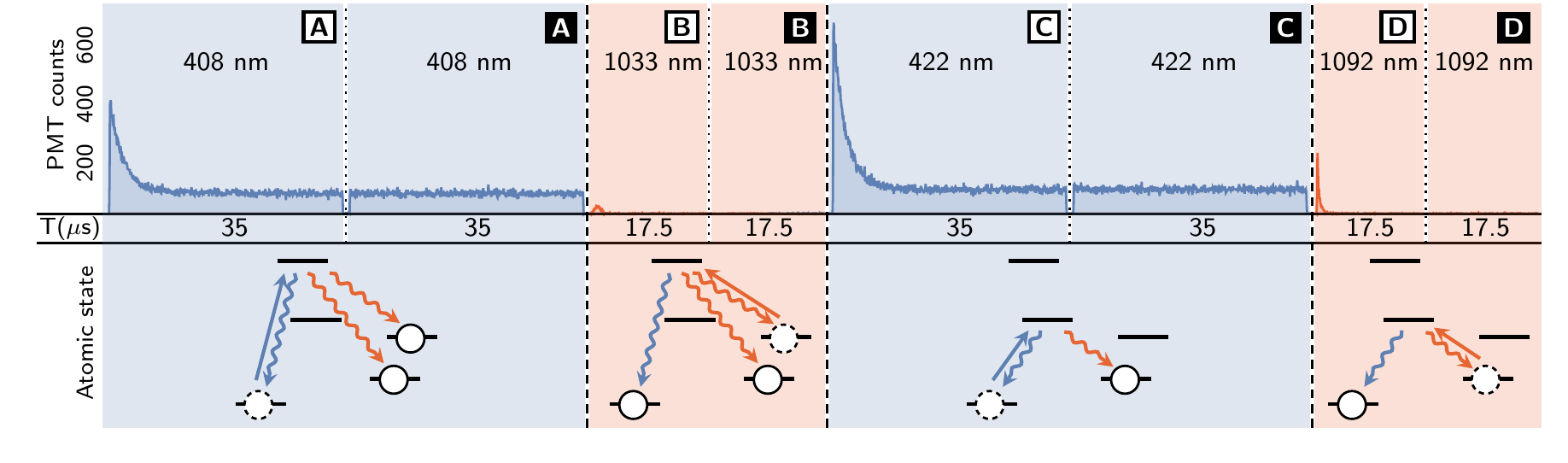}
\caption{Time-resolved fluorescence collected from the ion at 422 and 408 nm after $8\times10^6$ cycles of the \pthree/ branching ratio data sequence. Counts in the background intervals (black squares) are subtracted from the data intervals (white squares) to obtain  fluorescence from the ion only for each step. The length $T$ of each interval is indicated, and the laser turn-off time is 1 $\mu$s between each interval. The state of the ion in each step is depicted with both states losing population (dashed circle) and gaining population (solid circle).}
\label{fig:sequence}
\end{figure*}

As with the \pone/ measurement scheme by Ramm et. al., our sequence of population transfers is insensitive to detection efficiencies and many experimental variables. The long-lived shelving states \dthree/ and \dfive/ allow for the length of the measurement to far exceed the timescale needed for population transfer, rendering the measurement independent of laser power and frequency fluctuations as well as ion heating. There are no coherence effects or dark resonances since only one laser is on at a time, so our method is also insensitive to micromotion and magnetic field fluctuations. This distinguishes our method from a proposed $P_{3/2}$ branching ratio measurement scheme \cite{pruttivarasin2014thesis}, which not only requires an extra laser for the \pthree/--\dthree/ transition but also that two lasers be alternately pulsed for each step to avoid dark resonances, making the measurement sequence significantly longer.

To demonstrate this method, we experimentally measure the $5P_{1/2}$ and $5P_{3/2}$ branching ratios in \sr/. We trap single $^{88}$Sr$^+$ ions using a surface electrode Paul trap fabricated by Sandia National Laboratories \cite{clark2013experimental}. RF and DC confining fields are set such that the axial secular frequency of the ion is 600 kHz, with radial frequencies in the 3-4 MHz range and a 15 degree tilt in the radial plane. A magnetic field of 5.4 Gauss is applied normal to the trap to lift the degeneracy of the Zeeman states. Fluorescence from the ion is collected along the same axis by an in-vacuum 0.42 NA aspheric lens (Edmunds 49-696) into a single photon resolution photomultiplier tube (PMT, Hamamatsu H10682-210) with a filter that only passes light between 408 and 422 nm (Semrock FF01-415/10-25). The PMT signal is counted by an FPGA with arrival time binned into 2 ns intervals. The overall detection efficiency of the setup is approximately $4\times 10^{-3}$ at both 422 nm and 408 nm. Dipole transitions of the ion are addressed using frequency-stabilized diode lasers. To execute the experimental sequence, we switch laser beams on and off using acousto-optic modulators (AOMs) driven by FPGA-controlled direct digital synthesizers.

Each branching ratio measurement cycle begins with 100 $\mu$s of Doppler cooling using 422 nm and 1092 nm lasers. Subsequently, we turn on only IR lasers for 20 $\mu$s to ensure the ion is in the \sone/ state, then perform the experimental sequence. We ran the \pone/ and \pthree/ branching ratio measurement sequences for $1.9\times10^8$ and $6.4\times10^7$ cycles respectively for a run time of 13 and 6 hours each. For each step within the experimental sequence, the laser is turned on twice: first the data interval where population transfer occurs, then the background interval that is subtracted from the data interval to obtain only fluorescence from the ion. For the \pone/ experiment, the 422 nm and 1092 nm intervals are 35 $\mu$s and 25 $\mu$s in length respectively for both data and background, with 1 $\mu$s between each interval, and the \pthree/ experimental sequence is depicted in Fig. \ref{fig:sequence}. To arrive at final values for the branching ratios, we carefully calibrated the systematic sources of error in our experiment, which are summarized in Table \ref{tab:branchingtable}.

\newcommand\Tstrut{\rule{0pt}{2.6ex}}
\newcommand\Bstrut{\rule[-1.2ex]{0pt}{0pt}} 
\newcolumntype{R}[1]{>{\RaggedLeft\arraybackslash}p{#1}}
\begin{table}[!b]
\footnotesize
\centering
\begin{tabularx}{\linewidth}{l r r r}
\hline
\hline

& \multicolumn{3}{c}{Fractional shift and uncertainty}\\
\renewcommand{\arraystretch}{1.0}Error source&$p/(1-p)$&$q/(1-q)$&$s/(1-s)$\Tstrut\Bstrut\\
\hline
Counting statistics &  $\pm 16[-4]$ & $\pm 38[-4]$ & $\pm 33[-4]$  \\
Polarization alignment & \dots & $\pm19[-4]$ & $\pm19[-4]$ \\
PMT dead time & $46 \pm 2 [-5]$ & $76 \pm 5[-5]$ & $-45 \pm 1 [-5]$ \\
Finite laser durations &  $13 \pm 9[-8]$ & $\pm 3[-6]$ & $-6 \pm 2[-6]$ \\
AOM extinction ratio & $\pm 7 [-7]$ & $ \pm 3[-6]$ & $\pm  1[-6]$\\
Finite $D$ state lifetime & $354 \pm 4[-8]$ & $ -40 \pm 2[-7]$ & $ -124 \pm 2[-7]$ \\
\pone/ branching ratio & \dots & $\pm 16[-4]$ & \dots \\
\hline
\renewcommand{\arraystretch}{1.5}Total & $5 \pm 16[-4]$ & $8 \pm 45[-4]$ & $5 \pm 38[-4]$\Tstrut\Bstrut\\
\hline\hline
\end{tabularx}
\caption{List of systematic sources of error for branching ratios of \pone/--\sone/, \pthree/--\sone/, and \pthree/--\dfive/ and their contributions to the overall shift and uncertainty. Powers of 10 are in brackets.}
\label{tab:branchingtable}
\end{table}

\begin{figure*}
\includegraphics[width=\linewidth]{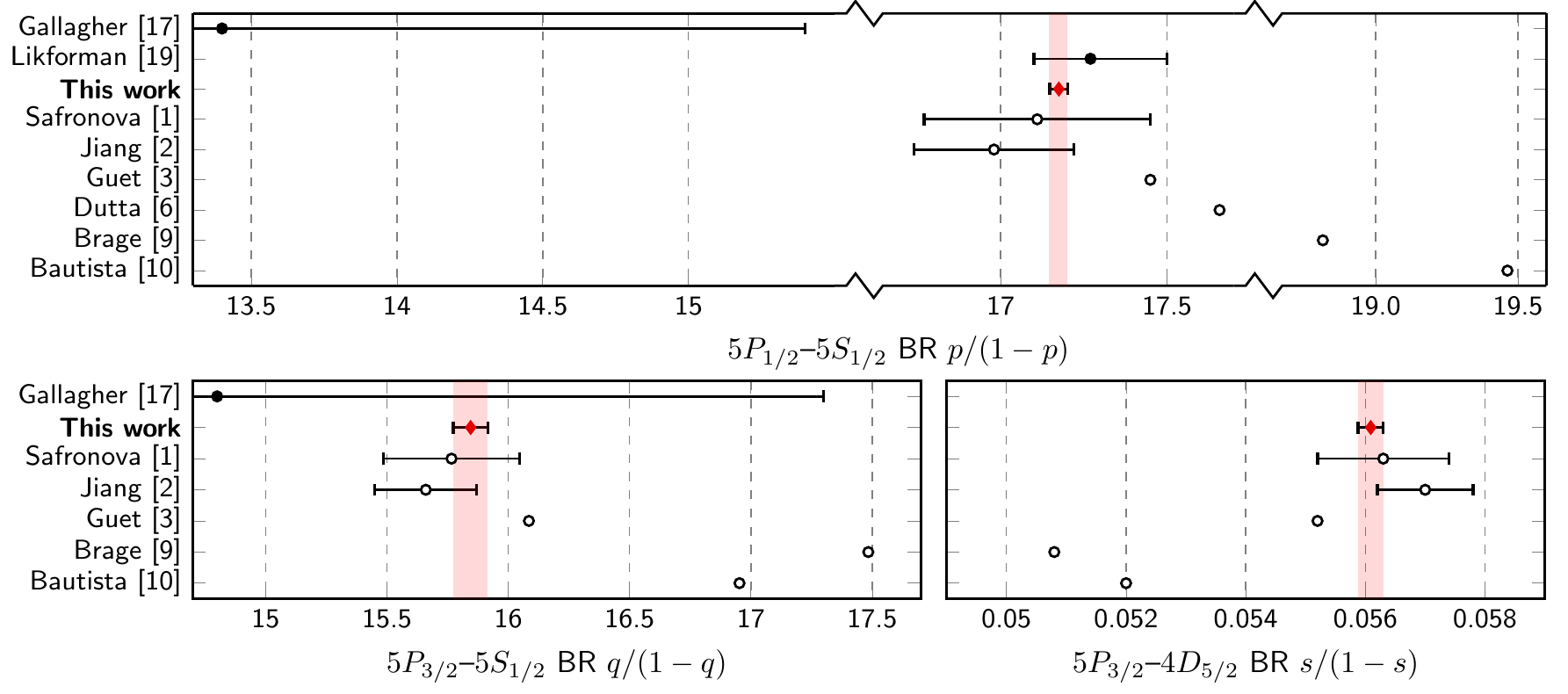}
\caption{Branching ratios (BR) for the \pone/ and \pthree/ excited states in Sr$^{+}$  obtained by this work (red diamond) and previous experimental (filled circle) and theoretical (empty circle) works. Error bars are included whenever uncertainties are provided by the source.}
\label{fig:comparison}
\end{figure*}

The polarization alignment error arises from the Hanle effect and is a function of the magnetic field, detector position, and  incident laser direction and polarization. In the \pone/ system, the $m=\pm1/2$ sublevels both emit radiation isotropically with 1:2 ratios of $\pi$- to $\sigma$-polarized light regardless of magnetic and electric fields, so this does not affect the measurement \cite{gallagher1963optical}. However, the Hanle effect is a major source of error for the \pthree/ system as the ratio of emitted $\pi$- to $\sigma$-polarized photons is 0:1 for $m=\pm 3/2$ sublevels and 2:1 for $m=\pm 1/2$ sublevels. The ratio of $\pi$ to $\sigma$ light emitted from Step A and Step B will therefore not be equal in general, biasing the fluorescence ratio.

To resolve this problem, we linearly polarize 408 and 1033 nm light to an axis 54.7$^\circ$ (the magic angle \cite{fano1973impact}) with respect to the magnetic field, which is set orthogonally to the laser beam. At the magic angle, the ratio of $\pi$- to $\sigma$- polarized light emitted during Steps A and B are both equal to 1:2. The difference between radiation patterns of $\pi$ and $\sigma$ photons and any birefringence effects in the detection system cancel out. We use a Glan-Taylor polarizer with $>$50 dB attenuation of the orthogonal polarization to align the 408 and 1033 nm laser polarizations to within 0.2 degrees of the magic angle. The error in aligning the laser polarization with respect to the magnetic field and setting the magnetic field to be orthogonal to the laser beam accounts for the polarization alignment error in Table \ref{tab:branchingtable}.

The effects of the other sources of systematics are accounted for by modeling the time-resolved fluorescence curves using optical Bloch equations to determine the shift and uncertainty contributed by each error source. PMT dead time, calibrated to be $20\pm1$ ns for our system using the method by Meeks and Siegel \cite{meeks2008dead}, leads to more undercounting at higher count rates. Finite laser durations reduces the fluorescence from the ion in each step in addition to preparing states imperfectly. The small amount of laser light still present when the AOMs are switched off (extinction ratios $>$60 dB) leads to slight coupling between undesirable states. The finite lifetime of the \dthree/ and \dfive/ states lead to extra counts in the blue intervals and reduced counts in the IR intervals. Off-resonant excitations, where the ion is excited to the wrong state by a collision or far-detuned laser, are found to contribute negligible errors to our system based on measuring the frequency of dark events while Doppler cooling the ion. We find that these sources of systematics do not limit our current level of accuracy. We also verify that the fluorescence from the ion is normally distributed when binned into 500,000 measurement cycles.

The largest source of error for both \pone/ and \pthree/ branching ratios is from counting statistics. This can be improved via either more measurement cycles, more ions, or greater collection efficiency, though for the latter two methods it is important to take into account the increased error from PMT dead time. Other errors can also be reduced via improvement of the experimental apparatus, such as more careful alignment of the laser polarization and using a PMT with less dead time. The only fundamental limitation to the accuracy of the technique is the uncertainty on the finite lifetimes of the \dthree/ and \dfive/ states, which restricts the length of the population inversion sequence, but the limit is many orders of magnitude below the current level of accuracy.

After applying systematic shifts and propagating uncertainties, we obtain for the \pone/ branching ratio $p/(1-p)=\p/$ and for the \pthree/ branching ratios $q/(1-q)=\q/$, $r/(1-r)=\rrr/$, and $s/(1-s)=\s/$, with errors representing 1$\sigma$ bounds. The corresponding branching fractions are $p=0.94498(8)$, $1-p=0.05502(8)$, $q=0.9406(2)$, $r=0.0063(3)$, and $s=0.0531(2)$.

The uncertainty of our results is at a level smaller than the discrepancy between previous experimental and theoretical results, as shown in Fig. \ref{fig:comparison}. Our value for the \pone/--\sone/ branching ratio agrees with the recent trapped ion experiment done by Likforman et. al. \cite{likforman2015precision} as well as theoretical values of Safronova \cite{safronova2010all} and Jiang et. al. \cite{jiang2009blackbody}, while disagreeing with the gas discharge chamber experiment by Gallagher \cite{gallagher1967oscillator}. For \pthree/, only the \pthree/--\sone/ branching ratio has been previously reported, also by Gallagher, which our value is in agreement with. We are also in agreement with theory values of Safronova and Jiang et. al. for all \pthree/ branching ratios. We note that Safronova's theoretical values have been found to be in good agreement with precision measurements of branching ratios and dipole matrix elements in other elements \cite{auchter2014measurement,hettrich2015measurement,ramm2013precision,safronova2011blackbody,safronova2011excitation}. We obtain a ten-fold improvement in accuracy for the \pone/ branching ratios over Likforman et. al. and a forty-fold improvement for the \pthree/--\dfive/ branching ratio over Gallagher.

For improving the precision of \sr/ atomic clocks, it is important to have accurate rates for transitions to the \dfive/ and \sone/ levels. Using the 6.63(7) ns \pthree/ lifetime value measured by Pinnington et. al. \cite{pinnington1995studies}, we obtain transition rates $A_{P_{3/2}-S_{1/2}}=1.425(15)\times 10^8$ s$^{-1}$ and $A_{P_{3/2}-D_{5/2}}=8.010(89)\times 10^6$ s$^{-1}$ using our measured branching fractions. These are significantly more accurate compared to previous best-known transition rates of $A_{P_{3/2}-S_{1/2}}=1.43(6)\times 10^8$ s$^{-1}$ and $A_{P_{3/2}-D_{5/2}}=8.7(1.5) \times 10^6$ s$^{-1}$ from Gallagher \cite{gallagher1967oscillator}. The uncertainty in transition rates is now dominated by the uncertainty in the \pthree/ lifetime, which must be reduced by at least an order of magnitude to contribute towards more accurate polarizabilities and blackbody radiation shifts for optical clocks \cite{jiang2009blackbody}.

In summary, we have introduced a novel method for measuring the branching ratio of the $J=3/2$ state in trapped ion systems and demonstrated its effectiveness by measurements in \sr/ at the sub-1\% level. Our scheme, as with the Ramm et. al. method for $J=1/2$ states, uses only dipole transition addressing lasers and is insensitive to detector efficiencies, laser and magnetic field fluctuations, as well as ion heating and micromotion. The branching ratios of many higher-up excited states, such as the $6S_{1/2}$ and $5D_{3/2}$ states in \sr/, can also be measured by applying this method of chaining measurement sequences of successive states with selective detection of dipole transitions. This scheme is also broadly applicable to excited states in other elements with a similar structure of decaying into a ground state and long-lived states, such as secondary clock standards $^{199}$Hg$^+$ and $^{171}$Yb$^+$, for which greater branching ratio and lifetime precision can reduce uncertainty from blackbody radiation as well \cite{oskay2006single,huntemann2016single}.

The authors acknowledge helpful discussions with Hartmut H\"{a}ffner, Christian Roos, and Luca Guidoni. This work was supported in part by the IARPA MQCO program and by the NSF Center for Ultracold Atoms.

\bibliographystyle{apsrev4-1}
\bibliography{references.bib}

\begin{thebibliography}{36}%
\makeatletter
\providecommand \@ifxundefined [1]{%
 \@ifx{#1\undefined}
}%
\providecommand \@ifnum [1]{%
 \ifnum #1\expandafter \@firstoftwo
 \else \expandafter \@secondoftwo
 \fi
}%
\providecommand \@ifx [1]{%
 \ifx #1\expandafter \@firstoftwo
 \else \expandafter \@secondoftwo
 \fi
}%
\providecommand \natexlab [1]{#1}%
\providecommand \enquote  [1]{``#1''}%
\providecommand \bibnamefont  [1]{#1}%
\providecommand \bibfnamefont [1]{#1}%
\providecommand \citenamefont [1]{#1}%
\providecommand \href@noop [0]{\@secondoftwo}%
\providecommand \href [0]{\begingroup \@sanitize@url \@href}%
\providecommand \@href[1]{\@@startlink{#1}\@@href}%
\providecommand \@@href[1]{\endgroup#1\@@endlink}%
\providecommand \@sanitize@url [0]{\catcode `\\12\catcode `\$12\catcode
  `\&12\catcode `\#12\catcode `\^12\catcode `\_12\catcode `\%12\relax}%
\providecommand \@@startlink[1]{}%
\providecommand \@@endlink[0]{}%
\providecommand \url  [0]{\begingroup\@sanitize@url \@url }%
\providecommand \@url [1]{\endgroup\@href {#1}{\urlprefix }}%
\providecommand \urlprefix  [0]{URL }%
\providecommand \Eprint [0]{\href }%
\providecommand \doibase [0]{http://dx.doi.org/}%
\providecommand \selectlanguage [0]{\@gobble}%
\providecommand \bibinfo  [0]{\@secondoftwo}%
\providecommand \bibfield  [0]{\@secondoftwo}%
\providecommand \translation [1]{[#1]}%
\providecommand \BibitemOpen [0]{}%
\providecommand \bibitemStop [0]{}%
\providecommand \bibitemNoStop [0]{.\EOS\space}%
\providecommand \EOS [0]{\spacefactor3000\relax}%
\providecommand \BibitemShut  [1]{\csname bibitem#1\endcsname}%
\let\auto@bib@innerbib\@empty
\bibitem [{\citenamefont {Safronova}(2010)}]{safronova2010all}%
  \BibitemOpen
  \bibfield  {author} {\bibinfo {author} {\bibfnamefont {U.~I.}\ \bibnamefont
  {Safronova}},\ }\href@noop {} {\bibfield  {journal} {\bibinfo  {journal}
  {Phys. Rev. A}\ }\textbf {\bibinfo {volume} {82}},\ \bibinfo {pages} {022504}
  (\bibinfo {year} {2010})}\BibitemShut {NoStop}%
\bibitem [{\citenamefont {Jiang}\ \emph {et~al.}(2009)\citenamefont {Jiang},
  \citenamefont {Arora}, \citenamefont {Safronova},\ and\ \citenamefont
  {Clark}}]{jiang2009blackbody}%
  \BibitemOpen
  \bibfield  {author} {\bibinfo {author} {\bibfnamefont {D.}~\bibnamefont
  {Jiang}}, \bibinfo {author} {\bibfnamefont {B.}~\bibnamefont {Arora}},
  \bibinfo {author} {\bibfnamefont {M.~S.}\ \bibnamefont {Safronova}}, \ and\
  \bibinfo {author} {\bibfnamefont {C.~W.}\ \bibnamefont {Clark}},\ }\href@noop
  {} {\bibfield  {journal} {\bibinfo  {journal} {J. Phys. B}\ }\textbf
  {\bibinfo {volume} {42}},\ \bibinfo {pages} {154020} (\bibinfo {year}
  {2009})}\BibitemShut {NoStop}%
\bibitem [{\citenamefont {Guet}\ and\ \citenamefont
  {Johnson}(1991)}]{guet1991relativistic}%
  \BibitemOpen
  \bibfield  {author} {\bibinfo {author} {\bibfnamefont {C.}~\bibnamefont
  {Guet}}\ and\ \bibinfo {author} {\bibfnamefont {W.~R.}\ \bibnamefont
  {Johnson}},\ }\href@noop {} {\bibfield  {journal} {\bibinfo  {journal} {Phys.
  Rev. A}\ }\textbf {\bibinfo {volume} {44}},\ \bibinfo {pages} {1531}
  (\bibinfo {year} {1991})}\BibitemShut {NoStop}%
\bibitem [{\citenamefont {Fortson}(1993)}]{fortson1993possibility}%
  \BibitemOpen
  \bibfield  {author} {\bibinfo {author} {\bibfnamefont {N.}~\bibnamefont
  {Fortson}},\ }\href@noop {} {\bibfield  {journal} {\bibinfo  {journal} {Phys.
  Rev. Lett.}\ }\textbf {\bibinfo {volume} {70}},\ \bibinfo {pages} {2383}
  (\bibinfo {year} {1993})}\BibitemShut {NoStop}%
\bibitem [{\citenamefont {Dzuba}\ and\ \citenamefont
  {Flambaum}(2011)}]{dzuba2011calculation}%
  \BibitemOpen
  \bibfield  {author} {\bibinfo {author} {\bibfnamefont {V.~A.}\ \bibnamefont
  {Dzuba}}\ and\ \bibinfo {author} {\bibfnamefont {V.~V.}\ \bibnamefont
  {Flambaum}},\ }\href@noop {} {\bibfield  {journal} {\bibinfo  {journal}
  {Phys. Rev. A}\ }\textbf {\bibinfo {volume} {83}},\ \bibinfo {pages} {052513}
  (\bibinfo {year} {2011})}\BibitemShut {NoStop}%
\bibitem [{\citenamefont {Dutta}\ and\ \citenamefont
  {Majumder}(2014)}]{dutta2014}%
  \BibitemOpen
  \bibfield  {author} {\bibinfo {author} {\bibfnamefont {N.~N.}\ \bibnamefont
  {Dutta}}\ and\ \bibinfo {author} {\bibfnamefont {S.}~\bibnamefont
  {Majumder}},\ }\href@noop {} {\bibfield  {journal} {\bibinfo  {journal}
  {Phys. Rev. A}\ }\textbf {\bibinfo {volume} {90}},\ \bibinfo {pages} {012522}
  (\bibinfo {year} {2014})}\BibitemShut {NoStop}%
\bibitem [{\citenamefont {Caffau}\ \emph {et~al.}(2016)\citenamefont {Caffau},
  \citenamefont {Andrievsky}, \citenamefont {Korotin}, \citenamefont {Origlia},
  \citenamefont {Oliva}, \citenamefont {Sanna}, \citenamefont {Ludwig},\ and\
  \citenamefont {Bonifacio}}]{caffau2016giano}%
  \BibitemOpen
  \bibfield  {author} {\bibinfo {author} {\bibfnamefont {E.}~\bibnamefont
  {Caffau}}, \bibinfo {author} {\bibfnamefont {S.}~\bibnamefont {Andrievsky}},
  \bibinfo {author} {\bibfnamefont {S.}~\bibnamefont {Korotin}}, \bibinfo
  {author} {\bibfnamefont {L.}~\bibnamefont {Origlia}}, \bibinfo {author}
  {\bibfnamefont {E.}~\bibnamefont {Oliva}}, \bibinfo {author} {\bibfnamefont
  {N.}~\bibnamefont {Sanna}}, \bibinfo {author} {\bibfnamefont {H.-G.}\
  \bibnamefont {Ludwig}}, \ and\ \bibinfo {author} {\bibfnamefont
  {P.}~\bibnamefont {Bonifacio}},\ }\href@noop {} {\bibfield  {journal}
  {\bibinfo  {journal} {Astron. Astrophys.}\ }\textbf {\bibinfo {volume}
  {585}},\ \bibinfo {pages} {A16} (\bibinfo {year} {2016})}\BibitemShut
  {NoStop}%
\bibitem [{\citenamefont {Hansen}\ \emph {et~al.}(2014)\citenamefont {Hansen},
  \citenamefont {Hansen}, \citenamefont {Christlieb}, \citenamefont {Yong},
  \citenamefont {Bessell}, \citenamefont {P{\'e}rez}, \citenamefont {Beers},
  \citenamefont {Placco}, \citenamefont {Frebel}, \citenamefont {Norris} \emph
  {et~al.}}]{hansen2014exploring}%
  \BibitemOpen
  \bibfield  {author} {\bibinfo {author} {\bibfnamefont {T.}~\bibnamefont
  {Hansen}}, \bibinfo {author} {\bibfnamefont {C.}~\bibnamefont {Hansen}},
  \bibinfo {author} {\bibfnamefont {N.}~\bibnamefont {Christlieb}}, \bibinfo
  {author} {\bibfnamefont {D.}~\bibnamefont {Yong}}, \bibinfo {author}
  {\bibfnamefont {M.}~\bibnamefont {Bessell}}, \bibinfo {author} {\bibfnamefont
  {A.~G.}\ \bibnamefont {P{\'e}rez}}, \bibinfo {author} {\bibfnamefont
  {T.}~\bibnamefont {Beers}}, \bibinfo {author} {\bibfnamefont
  {V.}~\bibnamefont {Placco}}, \bibinfo {author} {\bibfnamefont
  {A.}~\bibnamefont {Frebel}}, \bibinfo {author} {\bibfnamefont
  {J.}~\bibnamefont {Norris}},  \emph {et~al.},\ }\href@noop {} {\bibfield
  {journal} {\bibinfo  {journal} {Astrophys. J.}\ }\textbf {\bibinfo {volume}
  {787}},\ \bibinfo {pages} {162} (\bibinfo {year} {2014})}\BibitemShut
  {NoStop}%
\bibitem [{\citenamefont {Brage}\ \emph {et~al.}(1998)\citenamefont {Brage},
  \citenamefont {Wahlgren}, \citenamefont {Johansson}, \citenamefont
  {Leckrone},\ and\ \citenamefont {Proffitt}}]{brage1998theoretical}%
  \BibitemOpen
  \bibfield  {author} {\bibinfo {author} {\bibfnamefont {T.}~\bibnamefont
  {Brage}}, \bibinfo {author} {\bibfnamefont {G.~M.}\ \bibnamefont {Wahlgren}},
  \bibinfo {author} {\bibfnamefont {S.~G.}\ \bibnamefont {Johansson}}, \bibinfo
  {author} {\bibfnamefont {D.~S.}\ \bibnamefont {Leckrone}}, \ and\ \bibinfo
  {author} {\bibfnamefont {C.~R.}\ \bibnamefont {Proffitt}},\ }\href@noop {}
  {\bibfield  {journal} {\bibinfo  {journal} {Astrophys. J.}\ }\textbf
  {\bibinfo {volume} {496}},\ \bibinfo {pages} {1051} (\bibinfo {year}
  {1998})}\BibitemShut {NoStop}%
\bibitem [{\citenamefont {Bautista}\ \emph {et~al.}(2002)\citenamefont
  {Bautista}, \citenamefont {Gull}, \citenamefont {Ishibashi}, \citenamefont
  {Hartman},\ and\ \citenamefont {Davidson}}]{bautista2002excitation}%
  \BibitemOpen
  \bibfield  {author} {\bibinfo {author} {\bibfnamefont {M.~A.}\ \bibnamefont
  {Bautista}}, \bibinfo {author} {\bibfnamefont {T.~R.}\ \bibnamefont {Gull}},
  \bibinfo {author} {\bibfnamefont {K.}~\bibnamefont {Ishibashi}}, \bibinfo
  {author} {\bibfnamefont {H.}~\bibnamefont {Hartman}}, \ and\ \bibinfo
  {author} {\bibfnamefont {K.}~\bibnamefont {Davidson}},\ }\href@noop {}
  {\bibfield  {journal} {\bibinfo  {journal} {Mon. Not. R. Astro. Soc.}\
  }\textbf {\bibinfo {volume} {331}},\ \bibinfo {pages} {875} (\bibinfo {year}
  {2002})}\BibitemShut {NoStop}%
\bibitem [{\citenamefont {Gratton}\ and\ \citenamefont
  {Sneden}(1994)}]{gratton1994abundances}%
  \BibitemOpen
  \bibfield  {author} {\bibinfo {author} {\bibfnamefont {R.~G.}\ \bibnamefont
  {Gratton}}\ and\ \bibinfo {author} {\bibfnamefont {C.}~\bibnamefont
  {Sneden}},\ }\href@noop {} {\bibfield  {journal} {\bibinfo  {journal}
  {Astron. Astrophys.}\ }\textbf {\bibinfo {volume} {287}},\ \bibinfo {pages}
  {927} (\bibinfo {year} {1994})}\BibitemShut {NoStop}%
\bibitem [{\citenamefont {Siqueira-Mello}\ \emph {et~al.}(2015)\citenamefont
  {Siqueira-Mello}, \citenamefont {Andrievsky}, \citenamefont {Barbuy},
  \citenamefont {Spite}, \citenamefont {Spite},\ and\ \citenamefont
  {Korotin}}]{siqueira2015high}%
  \BibitemOpen
  \bibfield  {author} {\bibinfo {author} {\bibfnamefont {C.}~\bibnamefont
  {Siqueira-Mello}}, \bibinfo {author} {\bibfnamefont {S.}~\bibnamefont
  {Andrievsky}}, \bibinfo {author} {\bibfnamefont {B.}~\bibnamefont {Barbuy}},
  \bibinfo {author} {\bibfnamefont {M.}~\bibnamefont {Spite}}, \bibinfo
  {author} {\bibfnamefont {F.}~\bibnamefont {Spite}}, \ and\ \bibinfo {author}
  {\bibfnamefont {S.}~\bibnamefont {Korotin}},\ }\href@noop {} {\bibfield
  {journal} {\bibinfo  {journal} {Astron. Astrophys.}\ }\textbf {\bibinfo
  {volume} {584}},\ \bibinfo {pages} {A86} (\bibinfo {year}
  {2015})}\BibitemShut {NoStop}%
\bibitem [{\citenamefont {Bergemann}\ \emph {et~al.}(2012)\citenamefont
  {Bergemann}, \citenamefont {Hansen}, \citenamefont {Bautista},\ and\
  \citenamefont {Ruchti}}]{bergemann2012nlte}%
  \BibitemOpen
  \bibfield  {author} {\bibinfo {author} {\bibfnamefont {M.}~\bibnamefont
  {Bergemann}}, \bibinfo {author} {\bibfnamefont {C.~J.}\ \bibnamefont
  {Hansen}}, \bibinfo {author} {\bibfnamefont {M.}~\bibnamefont {Bautista}}, \
  and\ \bibinfo {author} {\bibfnamefont {G.}~\bibnamefont {Ruchti}},\
  }\href@noop {} {\bibfield  {journal} {\bibinfo  {journal} {Astron.
  Astrophys.}\ }\textbf {\bibinfo {volume} {546}},\ \bibinfo {pages} {A90}
  (\bibinfo {year} {2012})}\BibitemShut {NoStop}%
\bibitem [{\citenamefont {Ludlow}\ \emph {et~al.}(2015)\citenamefont {Ludlow},
  \citenamefont {Boyd}, \citenamefont {Ye}, \citenamefont {Peik},\ and\
  \citenamefont {Schmidt}}]{ludlow2015optical}%
  \BibitemOpen
  \bibfield  {author} {\bibinfo {author} {\bibfnamefont {A.~D.}\ \bibnamefont
  {Ludlow}}, \bibinfo {author} {\bibfnamefont {M.~M.}\ \bibnamefont {Boyd}},
  \bibinfo {author} {\bibfnamefont {J.}~\bibnamefont {Ye}}, \bibinfo {author}
  {\bibfnamefont {E.}~\bibnamefont {Peik}}, \ and\ \bibinfo {author}
  {\bibfnamefont {P.~O.}\ \bibnamefont {Schmidt}},\ }\href@noop {} {\bibfield
  {journal} {\bibinfo  {journal} {Rev. Mod. Phys.}\ }\textbf {\bibinfo {volume}
  {87}},\ \bibinfo {pages} {637} (\bibinfo {year} {2015})}\BibitemShut
  {NoStop}%
\bibitem [{\citenamefont {Madej}\ \emph {et~al.}(2012)\citenamefont {Madej},
  \citenamefont {Dub{\'e}}, \citenamefont {Zhou}, \citenamefont {Bernard},\
  and\ \citenamefont {Gertsvolf}}]{madej2012sr+}%
  \BibitemOpen
  \bibfield  {author} {\bibinfo {author} {\bibfnamefont {A.~A.}\ \bibnamefont
  {Madej}}, \bibinfo {author} {\bibfnamefont {P.}~\bibnamefont {Dub{\'e}}},
  \bibinfo {author} {\bibfnamefont {Z.}~\bibnamefont {Zhou}}, \bibinfo {author}
  {\bibfnamefont {J.~E.}\ \bibnamefont {Bernard}}, \ and\ \bibinfo {author}
  {\bibfnamefont {M.}~\bibnamefont {Gertsvolf}},\ }\href@noop {} {\bibfield
  {journal} {\bibinfo  {journal} {Phys. Rev. Lett.}\ }\textbf {\bibinfo
  {volume} {109}},\ \bibinfo {pages} {203002} (\bibinfo {year}
  {2012})}\BibitemShut {NoStop}%
\bibitem [{\citenamefont {Weyers}\ \emph {et~al.}(2011)\citenamefont {Weyers},
  \citenamefont {Gerginov}, \citenamefont {Nemitz}, \citenamefont {Li},\ and\
  \citenamefont {Gibble}}]{weyers2011distributed}%
  \BibitemOpen
  \bibfield  {author} {\bibinfo {author} {\bibfnamefont {S.}~\bibnamefont
  {Weyers}}, \bibinfo {author} {\bibfnamefont {V.}~\bibnamefont {Gerginov}},
  \bibinfo {author} {\bibfnamefont {N.}~\bibnamefont {Nemitz}}, \bibinfo
  {author} {\bibfnamefont {R.}~\bibnamefont {Li}}, \ and\ \bibinfo {author}
  {\bibfnamefont {K.}~\bibnamefont {Gibble}},\ }\href@noop {} {\bibfield
  {journal} {\bibinfo  {journal} {Metrologia}\ }\textbf {\bibinfo {volume}
  {49}},\ \bibinfo {pages} {82} (\bibinfo {year} {2011})}\BibitemShut {NoStop}%
\bibitem [{\citenamefont {Gallagher}(1967)}]{gallagher1967oscillator}%
  \BibitemOpen
  \bibfield  {author} {\bibinfo {author} {\bibfnamefont {A.}~\bibnamefont
  {Gallagher}},\ }\href@noop {} {\bibfield  {journal} {\bibinfo  {journal}
  {Phys. Rev.}\ }\textbf {\bibinfo {volume} {157}},\ \bibinfo {pages} {24}
  (\bibinfo {year} {1967})}\BibitemShut {NoStop}%
\bibitem [{\citenamefont {Ramm}\ \emph {et~al.}(2013)\citenamefont {Ramm},
  \citenamefont {Pruttivarasin}, \citenamefont {Kokish}, \citenamefont
  {Talukdar},\ and\ \citenamefont {H{\"a}ffner}}]{ramm2013precision}%
  \BibitemOpen
  \bibfield  {author} {\bibinfo {author} {\bibfnamefont {M.}~\bibnamefont
  {Ramm}}, \bibinfo {author} {\bibfnamefont {T.}~\bibnamefont {Pruttivarasin}},
  \bibinfo {author} {\bibfnamefont {M.}~\bibnamefont {Kokish}}, \bibinfo
  {author} {\bibfnamefont {I.}~\bibnamefont {Talukdar}}, \ and\ \bibinfo
  {author} {\bibfnamefont {H.}~\bibnamefont {H{\"a}ffner}},\ }\href@noop {}
  {\bibfield  {journal} {\bibinfo  {journal} {Phys. Rev. Lett.}\ }\textbf
  {\bibinfo {volume} {111}},\ \bibinfo {pages} {023004} (\bibinfo {year}
  {2013})}\BibitemShut {NoStop}%
\bibitem [{\citenamefont {Likforman}\ \emph {et~al.}(2015)\citenamefont
  {Likforman}, \citenamefont {Tugay{\'e}}, \citenamefont {Guibal},\ and\
  \citenamefont {Guidoni}}]{likforman2015precision}%
  \BibitemOpen
  \bibfield  {author} {\bibinfo {author} {\bibfnamefont {J.-P.}\ \bibnamefont
  {Likforman}}, \bibinfo {author} {\bibfnamefont {V.}~\bibnamefont
  {Tugay{\'e}}}, \bibinfo {author} {\bibfnamefont {S.}~\bibnamefont {Guibal}},
  \ and\ \bibinfo {author} {\bibfnamefont {L.}~\bibnamefont {Guidoni}},\
  }\href@noop {} {\bibfield  {journal} {\bibinfo  {journal} {arXiv:1511.07686}\
  } (\bibinfo {year} {2015})}\BibitemShut {NoStop}%
\bibitem [{\citenamefont {Gerritsma}\ \emph {et~al.}(2008)\citenamefont
  {Gerritsma}, \citenamefont {Kirchmair}, \citenamefont {Z{\"a}hringer},
  \citenamefont {Benhelm}, \citenamefont {Blatt},\ and\ \citenamefont
  {Roos}}]{gerritsma2008precision}%
  \BibitemOpen
  \bibfield  {author} {\bibinfo {author} {\bibfnamefont {R.}~\bibnamefont
  {Gerritsma}}, \bibinfo {author} {\bibfnamefont {G.}~\bibnamefont
  {Kirchmair}}, \bibinfo {author} {\bibfnamefont {F.}~\bibnamefont
  {Z{\"a}hringer}}, \bibinfo {author} {\bibfnamefont {J.}~\bibnamefont
  {Benhelm}}, \bibinfo {author} {\bibfnamefont {R.}~\bibnamefont {Blatt}}, \
  and\ \bibinfo {author} {\bibfnamefont {C.}~\bibnamefont {Roos}},\ }\href@noop
  {} {\bibfield  {journal} {\bibinfo  {journal} {Eur. Phys. J. D}\ }\textbf
  {\bibinfo {volume} {50}},\ \bibinfo {pages} {13} (\bibinfo {year}
  {2008})}\BibitemShut {NoStop}%
\bibitem [{\citenamefont {De~Munshi}\ \emph {et~al.}(2015)\citenamefont
  {De~Munshi}, \citenamefont {Dutta}, \citenamefont {Rebhi},\ and\
  \citenamefont {Mukherjee}}]{de2015precision}%
  \BibitemOpen
  \bibfield  {author} {\bibinfo {author} {\bibfnamefont {D.}~\bibnamefont
  {De~Munshi}}, \bibinfo {author} {\bibfnamefont {T.}~\bibnamefont {Dutta}},
  \bibinfo {author} {\bibfnamefont {R.}~\bibnamefont {Rebhi}}, \ and\ \bibinfo
  {author} {\bibfnamefont {M.}~\bibnamefont {Mukherjee}},\ }\href@noop {}
  {\bibfield  {journal} {\bibinfo  {journal} {Phys. Rev. A}\ }\textbf {\bibinfo
  {volume} {91}},\ \bibinfo {pages} {040501} (\bibinfo {year}
  {2015})}\BibitemShut {NoStop}%
\bibitem [{\citenamefont {Hettrich}\ \emph {et~al.}(2015)\citenamefont
  {Hettrich}, \citenamefont {Ruster}, \citenamefont {Kaufmann}, \citenamefont
  {Roos}, \citenamefont {Schmiegelow}, \citenamefont {Schmidt-Kaler},\ and\
  \citenamefont {Poschinger}}]{hettrich2015measurement}%
  \BibitemOpen
  \bibfield  {author} {\bibinfo {author} {\bibfnamefont {M.}~\bibnamefont
  {Hettrich}}, \bibinfo {author} {\bibfnamefont {T.}~\bibnamefont {Ruster}},
  \bibinfo {author} {\bibfnamefont {H.}~\bibnamefont {Kaufmann}}, \bibinfo
  {author} {\bibfnamefont {C.~F.}\ \bibnamefont {Roos}}, \bibinfo {author}
  {\bibfnamefont {C.~T.}\ \bibnamefont {Schmiegelow}}, \bibinfo {author}
  {\bibfnamefont {F.}~\bibnamefont {Schmidt-Kaler}}, \ and\ \bibinfo {author}
  {\bibfnamefont {U.~G.}\ \bibnamefont {Poschinger}},\ }\href@noop {}
  {\bibfield  {journal} {\bibinfo  {journal} {Phys. Rev. Lett.}\ }\textbf
  {\bibinfo {volume} {115}},\ \bibinfo {pages} {143003} (\bibinfo {year}
  {2015})}\BibitemShut {NoStop}%
\bibitem [{\citenamefont {Monz}\ \emph {et~al.}(2016)\citenamefont {Monz},
  \citenamefont {Nigg}, \citenamefont {Martinez}, \citenamefont {Brandl},
  \citenamefont {Schindler}, \citenamefont {Rines}, \citenamefont {Wang},
  \citenamefont {Chuang},\ and\ \citenamefont {Blatt}}]{monz2015realization}%
  \BibitemOpen
  \bibfield  {author} {\bibinfo {author} {\bibfnamefont {T.}~\bibnamefont
  {Monz}}, \bibinfo {author} {\bibfnamefont {D.}~\bibnamefont {Nigg}}, \bibinfo
  {author} {\bibfnamefont {E.~A.}\ \bibnamefont {Martinez}}, \bibinfo {author}
  {\bibfnamefont {M.~F.}\ \bibnamefont {Brandl}}, \bibinfo {author}
  {\bibfnamefont {P.}~\bibnamefont {Schindler}}, \bibinfo {author}
  {\bibfnamefont {R.}~\bibnamefont {Rines}}, \bibinfo {author} {\bibfnamefont
  {S.~X.}\ \bibnamefont {Wang}}, \bibinfo {author} {\bibfnamefont {I.~L.}\
  \bibnamefont {Chuang}}, \ and\ \bibinfo {author} {\bibfnamefont
  {R.}~\bibnamefont {Blatt}},\ }\href@noop {} {\bibfield  {journal} {\bibinfo
  {journal} {Science}\ }\textbf {\bibinfo {volume} {351}},\ \bibinfo {pages}
  {1068} (\bibinfo {year} {2016})}\BibitemShut {NoStop}%
\bibitem [{\citenamefont {Kurz}\ \emph {et~al.}(2008)\citenamefont {Kurz},
  \citenamefont {Dietrich}, \citenamefont {Shu}, \citenamefont {Bowler},
  \citenamefont {Salacka}, \citenamefont {Mirgon},\ and\ \citenamefont
  {Blinov}}]{kurz2008measurement}%
  \BibitemOpen
  \bibfield  {author} {\bibinfo {author} {\bibfnamefont {N.}~\bibnamefont
  {Kurz}}, \bibinfo {author} {\bibfnamefont {M.~R.}\ \bibnamefont {Dietrich}},
  \bibinfo {author} {\bibfnamefont {G.}~\bibnamefont {Shu}}, \bibinfo {author}
  {\bibfnamefont {R.}~\bibnamefont {Bowler}}, \bibinfo {author} {\bibfnamefont
  {J.}~\bibnamefont {Salacka}}, \bibinfo {author} {\bibfnamefont
  {V.}~\bibnamefont {Mirgon}}, \ and\ \bibinfo {author} {\bibfnamefont {B.~B.}\
  \bibnamefont {Blinov}},\ }\href@noop {} {\bibfield  {journal} {\bibinfo
  {journal} {Phys. Rev. A}\ }\textbf {\bibinfo {volume} {77}},\ \bibinfo
  {pages} {060501} (\bibinfo {year} {2008})}\BibitemShut {NoStop}%
\bibitem [{\citenamefont {Sahoo}\ \emph {et~al.}(2006)\citenamefont {Sahoo},
  \citenamefont {Islam}, \citenamefont {Das}, \citenamefont {Chaudhuri},\ and\
  \citenamefont {Mukherjee}}]{sahoo2006lifetimes}%
  \BibitemOpen
  \bibfield  {author} {\bibinfo {author} {\bibfnamefont {B.~K.}\ \bibnamefont
  {Sahoo}}, \bibinfo {author} {\bibfnamefont {M.~R.}\ \bibnamefont {Islam}},
  \bibinfo {author} {\bibfnamefont {B.~P.}\ \bibnamefont {Das}}, \bibinfo
  {author} {\bibfnamefont {R.~K.}\ \bibnamefont {Chaudhuri}}, \ and\ \bibinfo
  {author} {\bibfnamefont {D.}~\bibnamefont {Mukherjee}},\ }\href@noop {}
  {\bibfield  {journal} {\bibinfo  {journal} {Phys. Rev. A}\ }\textbf {\bibinfo
  {volume} {74}},\ \bibinfo {pages} {062504} (\bibinfo {year}
  {2006})}\BibitemShut {NoStop}%
\bibitem [{\citenamefont {Pruttivarasin}(2014)}]{pruttivarasin2014thesis}%
  \BibitemOpen
  \bibfield  {author} {\bibinfo {author} {\bibfnamefont {T.}~\bibnamefont
  {Pruttivarasin}},\ }\href@noop {} {Ph.D. thesis},\ \bibinfo  {school}
  {University of California, Berkeley} (\bibinfo {year} {2014})\BibitemShut
  {NoStop}%
\bibitem [{\citenamefont {Clark}\ \emph {et~al.}(2013)\citenamefont {Clark},
  \citenamefont {Blain}, \citenamefont {Benito}, \citenamefont {Chou},
  \citenamefont {Descour}, \citenamefont {Ellis}, \citenamefont {Haltli},
  \citenamefont {Heller}, \citenamefont {Kemme}, \citenamefont {Sterk},
  \citenamefont {Tabakov}, \citenamefont {Tigges}, \citenamefont {Maunz},\ and\
  \citenamefont {Stick}}]{clark2013experimental}%
  \BibitemOpen
  \bibfield  {author} {\bibinfo {author} {\bibfnamefont {C.}~\bibnamefont
  {Clark}}, \bibinfo {author} {\bibfnamefont {M.}~\bibnamefont {Blain}},
  \bibinfo {author} {\bibfnamefont {F.}~\bibnamefont {Benito}}, \bibinfo
  {author} {\bibfnamefont {C.-W.}\ \bibnamefont {Chou}}, \bibinfo {author}
  {\bibfnamefont {M.}~\bibnamefont {Descour}}, \bibinfo {author} {\bibfnamefont
  {R.}~\bibnamefont {Ellis}}, \bibinfo {author} {\bibfnamefont
  {R.}~\bibnamefont {Haltli}}, \bibinfo {author} {\bibfnamefont
  {E.}~\bibnamefont {Heller}}, \bibinfo {author} {\bibfnamefont
  {S.}~\bibnamefont {Kemme}}, \bibinfo {author} {\bibfnamefont
  {J.}~\bibnamefont {Sterk}}, \bibinfo {author} {\bibfnamefont
  {B.}~\bibnamefont {Tabakov}}, \bibinfo {author} {\bibfnamefont
  {C.}~\bibnamefont {Tigges}}, \bibinfo {author} {\bibfnamefont
  {P.}~\bibnamefont {Maunz}}, \ and\ \bibinfo {author} {\bibfnamefont
  {D.}~\bibnamefont {Stick}},\ }\href@noop {} {\bibfield  {journal} {\bibinfo
  {journal} {Bull. Am. Phys. Soc.}\ }\textbf {\bibinfo {volume} {58}},\
  \bibinfo {pages} {6} (\bibinfo {year} {2013})}\BibitemShut {NoStop}%
\bibitem [{\citenamefont {Gallagher}\ and\ \citenamefont
  {Lurio}(1963)}]{gallagher1963optical}%
  \BibitemOpen
  \bibfield  {author} {\bibinfo {author} {\bibfnamefont {A.}~\bibnamefont
  {Gallagher}}\ and\ \bibinfo {author} {\bibfnamefont {A.}~\bibnamefont
  {Lurio}},\ }\href@noop {} {\bibfield  {journal} {\bibinfo  {journal} {Phys.
  Rev. Lett.}\ }\textbf {\bibinfo {volume} {10}},\ \bibinfo {pages} {25}
  (\bibinfo {year} {1963})}\BibitemShut {NoStop}%
\bibitem [{\citenamefont {Fano}\ and\ \citenamefont
  {Macek}(1973)}]{fano1973impact}%
  \BibitemOpen
  \bibfield  {author} {\bibinfo {author} {\bibfnamefont {U.}~\bibnamefont
  {Fano}}\ and\ \bibinfo {author} {\bibfnamefont {J.~H.}\ \bibnamefont
  {Macek}},\ }\href@noop {} {\bibfield  {journal} {\bibinfo  {journal} {Rev.
  Mod. Phys.}\ }\textbf {\bibinfo {volume} {45}},\ \bibinfo {pages} {553}
  (\bibinfo {year} {1973})}\BibitemShut {NoStop}%
\bibitem [{\citenamefont {Meeks}\ and\ \citenamefont
  {Siegel}(2008)}]{meeks2008dead}%
  \BibitemOpen
  \bibfield  {author} {\bibinfo {author} {\bibfnamefont {C.}~\bibnamefont
  {Meeks}}\ and\ \bibinfo {author} {\bibfnamefont {P.~B.}\ \bibnamefont
  {Siegel}},\ }\href@noop {} {\bibfield  {journal} {\bibinfo  {journal} {Am. J.
  Phys.}\ }\textbf {\bibinfo {volume} {76}},\ \bibinfo {pages} {589} (\bibinfo
  {year} {2008})}\BibitemShut {NoStop}%
\bibitem [{\citenamefont {Auchter}\ \emph {et~al.}(2014)\citenamefont
  {Auchter}, \citenamefont {Noel}, \citenamefont {Hoffman}, \citenamefont
  {Williams},\ and\ \citenamefont {Blinov}}]{auchter2014measurement}%
  \BibitemOpen
  \bibfield  {author} {\bibinfo {author} {\bibfnamefont {C.}~\bibnamefont
  {Auchter}}, \bibinfo {author} {\bibfnamefont {T.~W.}\ \bibnamefont {Noel}},
  \bibinfo {author} {\bibfnamefont {M.~R.}\ \bibnamefont {Hoffman}}, \bibinfo
  {author} {\bibfnamefont {S.~R.}\ \bibnamefont {Williams}}, \ and\ \bibinfo
  {author} {\bibfnamefont {B.~B.}\ \bibnamefont {Blinov}},\ }\href@noop {}
  {\bibfield  {journal} {\bibinfo  {journal} {Phys. Rev. A}\ }\textbf {\bibinfo
  {volume} {90}},\ \bibinfo {pages} {060501} (\bibinfo {year}
  {2014})}\BibitemShut {NoStop}%
\bibitem [{\citenamefont {Safronova}\ and\ \citenamefont
  {Safronova}(2011{\natexlab{a}})}]{safronova2011blackbody}%
  \BibitemOpen
  \bibfield  {author} {\bibinfo {author} {\bibfnamefont {M.~S.}\ \bibnamefont
  {Safronova}}\ and\ \bibinfo {author} {\bibfnamefont {U.~I.}\ \bibnamefont
  {Safronova}},\ }\href@noop {} {\bibfield  {journal} {\bibinfo  {journal}
  {Phys. Rev. A}\ }\textbf {\bibinfo {volume} {83}},\ \bibinfo {pages} {012503}
  (\bibinfo {year} {2011}{\natexlab{a}})}\BibitemShut {NoStop}%
\bibitem [{\citenamefont {Safronova}\ and\ \citenamefont
  {Safronova}(2011{\natexlab{b}})}]{safronova2011excitation}%
  \BibitemOpen
  \bibfield  {author} {\bibinfo {author} {\bibfnamefont {U.~I.}\ \bibnamefont
  {Safronova}}\ and\ \bibinfo {author} {\bibfnamefont {M.~S.}\ \bibnamefont
  {Safronova}},\ }\href@noop {} {\bibfield  {journal} {\bibinfo  {journal}
  {Can. J. Phys.}\ }\textbf {\bibinfo {volume} {89}},\ \bibinfo {pages} {465}
  (\bibinfo {year} {2011}{\natexlab{b}})}\BibitemShut {NoStop}%
\bibitem [{\citenamefont {Pinnington}\ \emph {et~al.}(1995)\citenamefont
  {Pinnington}, \citenamefont {Berends},\ and\ \citenamefont
  {Lumsden}}]{pinnington1995studies}%
  \BibitemOpen
  \bibfield  {author} {\bibinfo {author} {\bibfnamefont {E.~H.}\ \bibnamefont
  {Pinnington}}, \bibinfo {author} {\bibfnamefont {R.~W.}\ \bibnamefont
  {Berends}}, \ and\ \bibinfo {author} {\bibfnamefont {M.}~\bibnamefont
  {Lumsden}},\ }\href@noop {} {\bibfield  {journal} {\bibinfo  {journal} {J.
  Phys. B}\ }\textbf {\bibinfo {volume} {28}},\ \bibinfo {pages} {2095}
  (\bibinfo {year} {1995})}\BibitemShut {NoStop}%
\bibitem [{\citenamefont {Oskay}\ \emph {et~al.}(2006)\citenamefont {Oskay},
  \citenamefont {Diddams}, \citenamefont {Donley}, \citenamefont {Fortier},
  \citenamefont {Heavner}, \citenamefont {Hollberg}, \citenamefont {Itano},
  \citenamefont {Jefferts}, \citenamefont {Delaney}, \citenamefont {Kim},
  \citenamefont {Levi}, \citenamefont {Parker},\ and\ \citenamefont
  {Bergquist}}]{oskay2006single}%
  \BibitemOpen
  \bibfield  {author} {\bibinfo {author} {\bibfnamefont {W.~H.}\ \bibnamefont
  {Oskay}}, \bibinfo {author} {\bibfnamefont {S.~A.}\ \bibnamefont {Diddams}},
  \bibinfo {author} {\bibfnamefont {E.~A.}\ \bibnamefont {Donley}}, \bibinfo
  {author} {\bibfnamefont {T.~M.}\ \bibnamefont {Fortier}}, \bibinfo {author}
  {\bibfnamefont {T.~P.}\ \bibnamefont {Heavner}}, \bibinfo {author}
  {\bibfnamefont {L.}~\bibnamefont {Hollberg}}, \bibinfo {author}
  {\bibfnamefont {W.~M.}\ \bibnamefont {Itano}}, \bibinfo {author}
  {\bibfnamefont {S.~R.}\ \bibnamefont {Jefferts}}, \bibinfo {author}
  {\bibfnamefont {M.~J.}\ \bibnamefont {Delaney}}, \bibinfo {author}
  {\bibfnamefont {K.}~\bibnamefont {Kim}}, \bibinfo {author} {\bibfnamefont
  {F.}~\bibnamefont {Levi}}, \bibinfo {author} {\bibfnamefont {T.~E.}\
  \bibnamefont {Parker}}, \ and\ \bibinfo {author} {\bibfnamefont {J.~C.}\
  \bibnamefont {Bergquist}},\ }\href@noop {} {\bibfield  {journal} {\bibinfo
  {journal} {Phys. Rev. Lett.}\ }\textbf {\bibinfo {volume} {97}},\ \bibinfo
  {pages} {020801} (\bibinfo {year} {2006})}\BibitemShut {NoStop}%
\bibitem [{\citenamefont {Huntemann}\ \emph {et~al.}(2016)\citenamefont
  {Huntemann}, \citenamefont {Sanner}, \citenamefont {Lipphardt}, \citenamefont
  {Tamm},\ and\ \citenamefont {Peik}}]{huntemann2016single}%
  \BibitemOpen
  \bibfield  {author} {\bibinfo {author} {\bibfnamefont {N.}~\bibnamefont
  {Huntemann}}, \bibinfo {author} {\bibfnamefont {C.}~\bibnamefont {Sanner}},
  \bibinfo {author} {\bibfnamefont {B.}~\bibnamefont {Lipphardt}}, \bibinfo
  {author} {\bibfnamefont {C.}~\bibnamefont {Tamm}}, \ and\ \bibinfo {author}
  {\bibfnamefont {E.}~\bibnamefont {Peik}},\ }\href@noop {} {\bibfield
  {journal} {\bibinfo  {journal} {Phys. Rev. Lett.}\ }\textbf {\bibinfo
  {volume} {116}},\ \bibinfo {pages} {063001} (\bibinfo {year}
  {2016})}\BibitemShut {NoStop}%
\end{thebibliography}%

\end{document}